\documentclass[10pt]{iopart}

\usepackage{graphicx}
\usepackage{subfigure}
\usepackage{verbatim}
\usepackage{dcolumn}% Align table columns on decimal point
\usepackage{bm}% bold math
\usepackage{color}
\usepackage[toc]{appendix}
\usepackage{stmaryrd}
\usepackage{amsthm}
\usepackage{hhline}
\usepackage{amssymb}
\usepackage{iopams}
\usepackage{cite}
\usepackage{tikz}
\usetikzlibrary{arrows,shapes,calc,matrix}

\expandafter\let\csname equation*\endcsname\relax
\expandafter\let\csname endequation*\endcsname\relax
\usepackage{amsmath}
\usepackage{xstring}

\makeatletter
\newcommand\footnoteref[1]{\protected@xdef\@thefnmark{\ref{#1}}\@footnotemark}
\makeatother

\newcommand\invisiblesection[1]{%
	\refstepcounter{section}%
	\addcontentsline{toc}{section}{\protect\numberline{\thesection}#1}%
	\sectionmark{#1}}

\newcommand{\pd}{\partial}

\newcommand{\ex}[1]{\exp{\left(#1\right)}}

\newcommand{\bla}{bla\\bla\\bla\\bla\\bla}
\usepackage{fmtcount}

\renewcommand{\appendix}{
}

\newcommand{\draftmode}{1}    %to control draft colors below
\newcommand{\notetoself}[1]{\ifnum \draftmode=1 {\color[rgb]{0,0,0.8} [#1]} \fi}  %notes to self in blue when \draftmode==1.  invisible otherwise
\newcommand{\cuttext}[1]{\ifnum \draftmode=1 {\color[rgb]{0,0.5,0} [#1]} \fi}  %cut out text in green when \draftmode==1.  invisible otherwise
\newcommand{\warntext}[1]{\ifnum \draftmode=1 {\color[rgb]{0.9,0.6,0} #1} \else {#1} \color{black} \fi}
\newcommand{\aref}[1]{{Appendix~\hyperref[#1]{A}}}
\newcommand{\bref}[1]{{Appendix~\hyperref[#1]{B}}}
\usepackage[colorlinks=true,citecolor=blue,linkcolor=blue,urlcolor=blue]{hyperref}
\usepackage{doi}
\usepackage{cleveref}
\begin{document}

\title{Quantum Otto engines at relativistic energies}

\author{Nathan M. Myers\textsuperscript{1,2,6}, Obinna Abah\textsuperscript{3,4}, and Sebastian Deffner\textsuperscript{1,5}}
\address{$^1$Department of Physics, University of Maryland, Baltimore County, Baltimore, MD 21250, USA}
\address{$^2$Computer, Computational and Statistical Sciences Division, Los Alamos National Laboratory, Los Alamos, New Mexico 87545, USA}
\address{$^3$Centre for Theoretical Atomic, Molecular, and Optical Physics, School of Mathematics and Physics, Queen's University Belfast, United Kingdom, BT7 1NN}
\address{$^4$Joint Quantum Centre (JQC) Durham-Newcastle, School of Mathematics, Statistics, and Physics, Newcastle University, Newcastle upon Tyne, NE1 7RU, United Kingdom}
\address{$^5$Instituto de F\'{i}sica `Gleb Wataghin', Universidade Estadual de Campinas, 13083-859, Campinas, S\~{a}o Paulo, Brazil}
\address{$^6$Author to whom any correspondence should be addressed.}
\ead{myersn1@umbc.edu}

\begin{abstract}
Relativistic quantum systems exhibit unique features not present at lower energies, such as the existence of both particles and antiparticles, and restrictions placed on the system dynamics due to the light cone. In order to understand what impact these relativistic phenomena have on the performance of quantum thermal machines we analyze a quantum Otto engine with a working medium of a relativistic particle in an oscillator potential evolving under Dirac or Klein-Gordon dynamics. We examine both the low-temperature, non-relativistic and high-temperature, relativistic limits of the dynamics and find that the relativistic engine operates with higher work output, but an effectively reduced compression ratio, leading to significantly smaller efficiency than its non-relativistic counterpart. Using the framework of endoreversible thermodynamics we determine the efficiency at maximum power of the relativistic engine, and find it to be equivalent to the Curzon-Ahlborn efficiency.
\end{abstract}

\section{Introduction} 
\label{sec:1}

Using relativistic phenomena as a source of power has long been a staple of popular culture, such as the matter-antimatter annihilation reactors that power the warp drives of \textit{Star Trek} \cite{Sternbach1991, Lentz2021, Bobrick2021}. However, with rapidly developing nanoscale experimental control and the recent discovery of ``Dirac materials", condensed matter systems including graphene \cite{Beenakker2008} and Weyl semimetals \cite{Yan2017}, with linear dispersion relations whose low-energy excitations behave like massless relativistic particles \cite{Wehling2014}, relativistic quantum engines are no longer only the realm of science fiction.      

Since the beginnings of thermodynamics, the study of heat engines has played an integral role in understanding the thermodynamic behavior of a wide variety of systems \cite{kondepudi}. The discovery by Scovil and Schulz-DuBois that a three-level maser could be modeled as a continuous heat engine \cite{Scovil1959} opened the door to using the framework of heat engines to extend the principles of thermodynamics to the quantum regime. 

Since then, the study of quantum heat engines has expanded to a massive range of different systems and implementations. Works have examined the role of coherence \cite{Scully2003, Scully2011, Uzdin2016, Watanabe2017, Dann2020, Feldmann2012, Hardal2015, Hammam2021}, quantum correlations \cite{Barrios2021}, many-body effects \cite{Hardal2015, Beau2016, Li2018, Chen2019, Watanabe2020}, quantum uncertainty \cite{Kerremans2021}, degeneracy \cite{Pena2017, Barrios2018}, endoreversible cycles \cite{Deffner2018, Smith2020, Myers2021}, finite-time cycles \cite{Cavina2017, Feldmann2012, Zheng2016, Raja2020}, energy optimization \cite{Singh2020}, shortcuts to adiabaticity \cite{Abah2017, Abah2018, Abah2019, Beau2016, Campo2014, Funo2019, Bonanca2019, Baris2019, Dann2020, Li2018}, efficiency and power statistics \cite{Denzler2020, Denzler20202, Denzler20203}, and comparisons between classical and quantum machines \cite{Quan2007, Gardas2015, Friedenberger2017, Deffner2018}. Implementations have been proposed in harmonically confined single ions \cite{Abah2012}, magnetic systems \cite{Pena2015}, atomic clouds \cite{Niedenzu2019}, transmon qubits \cite{Cherubim2019}, optomechanical systems \cite{Zhang2014, Dechant2015}, and quantum dots \cite{Pena2019, Pena2020}. Quantum heat engines have been experimentally implemented using nanobeam oscillators \cite{Klaers2017}, atomic collisions \cite{Bouton2021}, and two-level ions \cite{Horne2020}. 

An area that has seen comparatively little exploration is the impact of special relativistic effects on quantum heat engine performance. Relativistic effects have been shown to significantly influence non-equilibrium thermodynamic behavior. In Dirac materials the strong spin-orbit coupling leads to modifications of the Boltzmann transport equations that accounts for additional contributions from the spin current \cite{Werner2019, Kashuba2020}. In the context of classical stochastic thermodynamics, it has been shown that the restrictions imposed by the light cone lead to modified behavior for heat \cite{Pal2020, Paraguassu2021}. The same has been shown to be true for the work distribution in relativistic quantum systems \cite{Deffner2015}. Recent works have examined quantum heat engine cycles for relativistic particles in a square well potential \cite{Munoz2012, Pena2016, Purwanto2016, Yin2018, Akbar2018, Setyo2018, Chattopadhyay2019}, and a non-relativistic working medium interacting with a relativistic bath \cite{Papadatos2021}. Heat engine implementations in Dirac materials such as graphene have also been proposed \cite{Pena2015, Fadaie2018, Mani2019, Pena2020}. 

Dirac materials are an area of intensive research within condensed matter physics, motivated in part by a wide range of potential applications. The high mobility of the charge carriers in Dirac materials opens up potential applications in high speed electronics \cite{Yan2017}, and their long spin lifetimes make them promising candidates for spintronic devices \cite{Pesin2012} and quantum memory systems \cite{Chen2021}. However, fully exploiting the advantages that such relativistic quantum systems provide necessitates a detailed understanding of their fundamental thermodynamic behavior. The analysis of heat engines provides us with a well-established and device-oriented framework for probing exactly this behavior.       

In this paper, we examine the performance of an endoreversible quantum Otto engine with a working medium of a single relativistic particle, either fermion or boson, in an oscillator potential. We demonstrate that the relativistic engine operates with an effectively reduced compression ratio due to restrictions on the dynamics from the light cone, resulting in lower efficiency in comparison to a non-relativistic working medium. We determine the efficiency at maximum power in the relativistic regime, and find it to be equivalent to the Curzon-Ahlborn efficiency, the efficiency at maximum power achieved by a classical Otto engine \cite{Leff1987}. In Section \ref{sec:2} we provide some necessary background on the Dirac and Klein-Gordon oscillators, relativistic extensions of the typical quantum harmonic oscillator. In Section \ref{sec:3} we examine the equilibrium thermodynamics of the single particle relativistic oscillators, including a derivation of the partition function and canonical distribution that takes into account both the particle and antiparticle solutions. In Section \ref{sec:4} we determine the efficiency, power, and efficiency at maximum power of an endoreversible quantum Otto cycle in both the high-temperature, relativistic and low-temperature, non-relativistic limits. Finally, in Section \ref{sec:5} we conclude with a discussion of possible experimental systems in which such a cycle could be implemented.             

\section{Relativistic Oscillators} 
\label{sec:2}

In the first quantization framework, the Dirac and Klein-Gordon equations provide the most well-known and pedagogical approach to relativistic quantum mechanics. The Dirac equation describes the dynamics of spin-1/2 fermions, while the Klein-Gordon equation applies to spinless bosons \cite{Baym1990}. The Dirac oscillator, first introduced by Moshinsky and Szczepaniak \cite{Moshinsky1989}, was derived by reverse engineering the Dirac equation Hamiltonian that would reduce to the familiar quantum harmonic oscillator in the non-relativistic limit. Moshinsky and Szczepaniak realized that, due to the first order nature of the Dirac equation, the Dirac Oscillator Hamiltonian would need to be linear in both coordinates and momentum. The corresponding $(1+1)$-dimensional, time independent Dirac equation reads \cite{Moshinsky1989, Decastro2003},
\begin{equation}
	\label{eq:DiracEqn}
	E \psi = \left[c \sigma_x \cdot \left(p - im \omega x \cdot \sigma_z \right)+mc^2\sigma_z\right]\psi,
\end{equation}
where $m$ is the particle mass, $\omega$ is the oscillator frequency, $\sigma_x$ and $\sigma_z$ are the corresponding Pauli spin matrices, and $\psi = \left[\psi_1 \,\, \psi_2\right]^{\top}$ is the bispinor.  

Equation \eqref{eq:DiracEqn} can be separated into two coupled differential equations,
\begin{equation}
	\label{eq:DiracPart1}
	(E - mc^2)\psi_1 = c \left(p + i m \omega x\right) \psi_2,
\end{equation} 
and,
\begin{equation}
	\label{eq:DiracPart2}
	(E + mc^2)\psi_2 = c \left(p - i m \omega x\right) \psi_1.
\end{equation} 
Combining Eqs. (\ref{eq:DiracPart1}) and (\ref{eq:DiracPart2}) we can eliminate $\psi_2$,
\begin{equation}
	\label{eq:KGOscillator}
	\left( \frac{p^2}{2m} + \frac{1}{2}m \omega^2 x^2 + \frac{1}{2} m c^2 - \frac{E^2}{2 m c^2} \right) \psi_1 = 0.
\end{equation}
We note that this is the familiar harmonic oscillator Schr\"{o}dinger equation with the energy replaced by the relativistic energy-momentum relation. As such, Eq. (\ref{eq:KGOscillator}) is identical to the Klein-Gordon equation for an oscillator potential \cite{Bruce1993}. The fact that the Dirac and Klein-Gordon equations coincide in this instance is a reflection of the fact that the oscillator potential does not interact with the spin, resulting in no spin-orbit coupling in one dimension \cite{Dominguez1990}. Consequently, the following analysis is equally applicable to both fermions and bosons.  

Equation (\ref{eq:KGOscillator}) is analytically solvable. The energy spectrum is \cite{Moshinsky1989, Dominguez1990, Bruce1993, Rao2007},
\begin{equation}
	\label{eq:DiracSpectrum}
	E_n = \pm \sqrt{2\left(n+\frac{1}{2}\right)\hbar \omega m c^2 + m^2 c^4},
\end{equation}
where $n = \{0, 1, 2, ...\}$. The positive branch corresponds to the particle energies and the negative branch to the anti-particle energies. Note that for the case of charge-neutral bosons the positive and negative energy solutions of the Klein-Gordon equation are completely symmetric \cite{Baym1990}. Subtracting off the rest energy, $mc^2$, we can rewrite the energy spectrum in terms of $\lambda = \hbar \omega / m c^2$, 
\begin{equation}
	\label{eq:energyCompton}
	\epsilon_n = \pm mc^2 \left(\sqrt{2 \lambda \left(n+\frac{1}{2}\right)+1}-1\right).
\end{equation}
In the limit of $\lambda \ll 1$, which corresponds to the ``classical" (still quantum, but non-relativistic) regime where $\hbar \omega$ is much smaller than the rest energy, the positive energy branch of Eq. (\ref{eq:energyCompton}) simplifies to,
\begin{equation}
	\epsilon_n = \hbar \omega \left(n+\frac{1}{2}\right),
\end{equation}
Thus we see that the energy spectrum of the one-dimensional Dirac oscillator reduces to that of the typical quantum harmonic oscillator. 

\section{Relativistic Equilibrium Thermodynamics}
\label{sec:3}

\subsection{Relativistic Canonical Ensemble}

The first step in understanding the relativistic thermodynamics of the Dirac oscillator is to determine the correct expression for the equilibrium distribution when the system is placed in contact with a thermal bath. Under Schr\"{o}dinger dynamics the equilibrium distribution for a quantum system weakly coupled to a large classical bath at inverse temperature $\beta$ is given by the Gibbs state \cite{Deffner2019book},
\begin{equation}
	\rho = \frac{1}{Z} \ex{- \beta H},
\end{equation} 
where $Z$ is the canonical partition function,
\begin{equation}
	Z = \mathrm{tr}\{\ex{-\beta H}\} = \sum_{n = 0}^{\infty} \ex{-\beta E_n}.
\end{equation}  
This definition presents issues for quantum systems evolving under Dirac or Klein-Gordon dynamics, as the partition function sum will diverge for the negative energy solutions. To determine the proper equilibrium state for relativistic quantum systems we repeat the standard canonical ensemble derivation \cite{Callen, Huang2009book, Deffner2016, Deffner2019book} while keeping careful account of both positive and negative energy states.

Consider a system of $k$ relativistic oscillators with $N$ energy quanta that can be distributed among them. The total energy for this system is given by,
\begin{equation}
	\label{eq:EnergySum}
	E = n_1^+ \epsilon_1 + n_2^+ \epsilon_2 + ... + n_k^+ \epsilon_k - n_1^- \epsilon_1 - n_2^- \epsilon_2 - ... - n_k^- \epsilon_k,
\end{equation}
where $n_i^+$ and $n_i^-$ are the occupation numbers of the $i$th positive or negative state with energy of magnitude $\epsilon_i$. 

\begin{figure}
	\centering
	\includegraphics[width=0.6\textwidth]{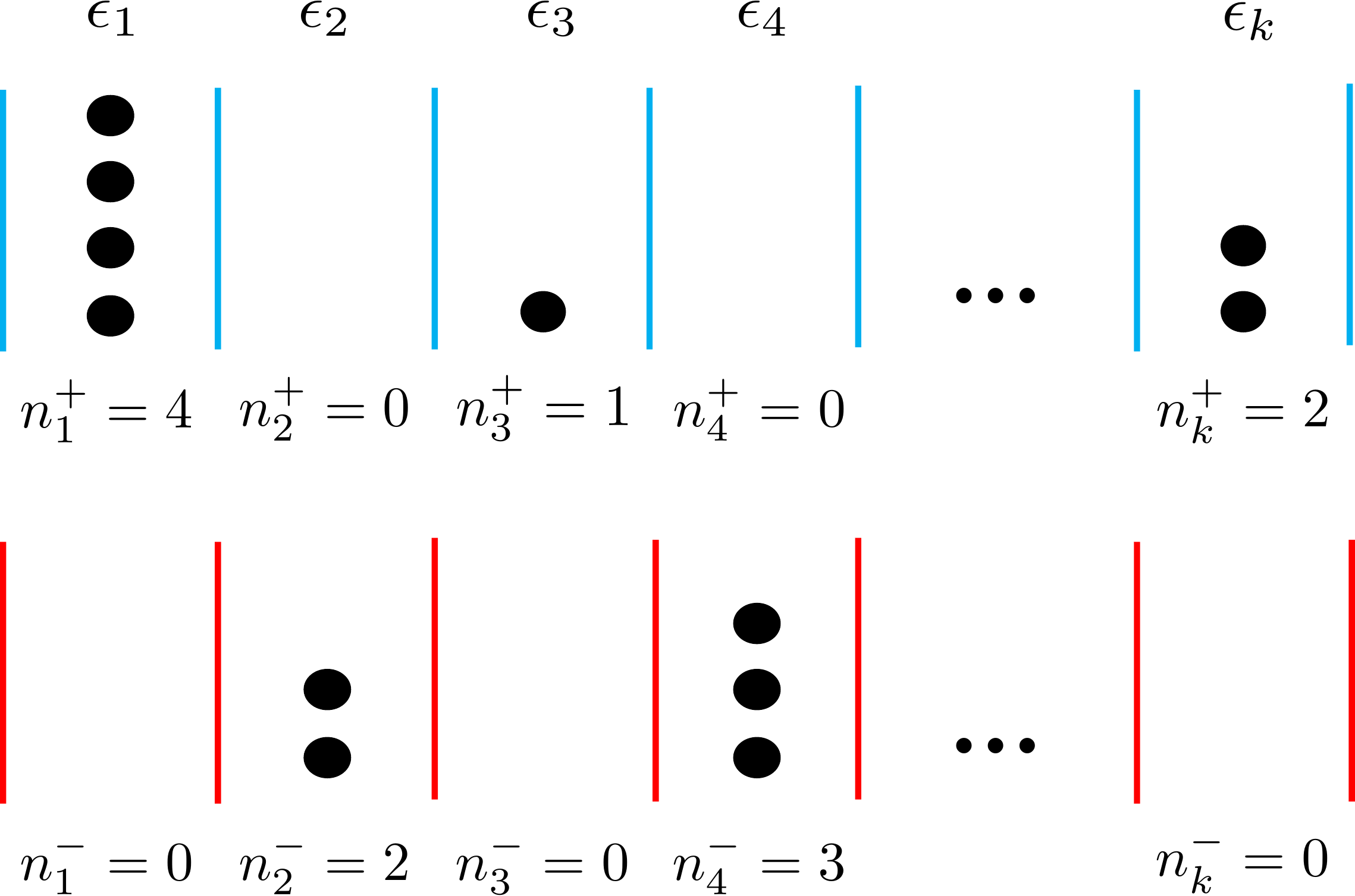}
	\caption{\label{fig:counting} Illustration of the combinatorics of distributing energy quanta across $k$ different relativistic oscillators. The top (blue partitions) and bottom (red partitions) rows represent the positive and negative energy solutions of magnitude $\epsilon_i$, respectively. Note that here we have accounted for the ``stability of the Dirac sea", which requires that the positive and negative energy states cannot be simultaneously occupied.}
\end{figure}

It is important to note that the Dirac oscillator Hamiltonian meets the conditions for a class of relativistic quantum systems that display a ``stability of the Dirac Sea," meaning that the positive and negative energy solutions do not mix \cite{Moreno1990, Martinez1991}. The stability of the Dirac sea can be derived by performing a Foldy-Wouthuysen canonical transformation on the Dirac Oscillator Hamiltonian. The transformed Hamiltonian can then be written in terms of only the even roots of the square of the Dirac Oscillator Hamiltonian \cite{Moreno1990}. As the transformation preserves the eigenvalues, this indicates that the positive and negative branches cannot be simultaneously occupied. This can also be interpreted as a supersymmetric partnering of the positive and negative energy solutions \cite{Moreno1990}. In the current context this means that either $n_i^+$ \textit{or} $n_i^-$ is non-zero, but not both. A graphical demonstration of the counting for distributing energy quanta across $k$ oscillators is shown in Fig. \ref{fig:counting}. The absence of mixed positive and negative energy states also neatly removes the issue of any fixed total energy having an infinite number of ``zero" energy states consisting of equal occupations in $n_i^+$ and $n_i^-$. With this in mind we define $n_i$ as the occupation of the $i$th state, either positive or negative. Equation (\ref{eq:EnergySum}) can then be expressed as,
\begin{equation}
	E = \sum_{i+}^{k^+} n_i \epsilon_i - \sum_{i-}^{k^-} n_i \epsilon_i,
\end{equation}
where $k^+$ and $k^-$ are the number of occupied positive and negative energy oscillators, respectively, with $k = k^+ + k^-$. The notation $i+$ ($i-$) indicates summations over the occupied positive (negative) energy states.   

Following the typical textbook approach \cite{Huang2009book}, we maximize the multiplicity of states subject to the constraints of fixed total energy and number of energy quanta. The full details of this derivation are provided in the \hyperref[Appendix]{Appendix}. Due to the stability of the Dirac Sea, we find that the positive and negative energy states can be treated separately, leading to the individual partition functions, 
\begin{equation}
	Z^+ = \sum_{i+} \ex{- \beta^+ \epsilon_i},
\end{equation}
for the positive energy solutions and, 
\begin{equation}
	\label{eq:NegPF}
	Z^- = \sum_{i-} \ex{\beta^- \epsilon_i},
\end{equation} 
for the negative energy solutions. The total partition function for the relativistic oscillator is then given by the product of the partition functions for the positive and negative energy solutions,
\begin{equation}
	Z = Z^+ Z^-.
\end{equation}
This factorization of the partition function is characteristic of systems consisting of a mixture of independent components \cite{Callen}. These results demonstrate that the equilibrium state of the Dirac oscillator can be treated as an ideal mixture of two gases, one composed of particles and the other of antiparticles. 

Recalling that $\epsilon_i$ are the energy \textit{magnitudes} it seems we have not resolved our original problem, as Eq. (\ref{eq:NegPF}) will still diverge. However, at this point we have not identified $\beta^+$ and $\beta^-$ with physical quantities, currently they are nothing more than Lagrange multipliers arising from the maximization of the state multiplicity. Following the typical steps \cite{Huang2009book, Deffner2019book} we see that $\beta^+$ gives the inverse temperature for the positive energy states and $\beta^-$ the negative inverse temperature for the negative energy states. As the equilibrium state is nothing more than the Gibbs state for a mixed ideal gas, we note that it also satisfies the Kubo-Martin-Schwinger conditions for an equilibrium state \cite{Bratteli1987book}. 

\subsection{Relativistic Thermodynamic Quantities}

With the proper equilibrium state and partition function determined, we can next turn to determining the equilibrium thermodynamic behavior of the Dirac Oscillator. The internal energy, free energy, entropy, and heat capacity can all be determined from the partition function \cite{Callen, Huang2009book}, 
\begin{equation}
	\begin{split}
		\label{eq:thermofunc}
		& E = - \frac{\partial }{\partial \beta} \ln(Z), \qquad \,
		F = - \frac{1}{\beta} \ln(Z), \\
		& S = k_{\mathrm{B}} \beta^2 \frac{\partial F}{\partial \beta}, \qquad \quad \,
		C = -k_{\mathrm{B}} \beta^2 \frac{\partial E}{\partial \beta}, 
	\end{split} 
\end{equation}
where $Z = Z^+Z^-$. 

Let us consider the positive energy states. Using Eq. (\ref{eq:energyCompton}) the partition function becomes,
\begin{equation}
	\label{eq:PartPlus}
	Z^+ = \sum_{n = 0}^{\infty} \exp{\left(- \beta mc^2 \left[\sqrt{2 \lambda (n+1)+1}-1\right]\right)}.
\end{equation}
We note that for relevant temperatures and frequencies, such as those consistent with a Dirac oscillator implemented in a trapped ion system \cite{Bermudez2007}, the spacing of the energy spectrum will be very small. Using experimentally achievable parameters \cite{Leibfried2003} for the relativistic regime corresponding to $\lambda \sim 10^8$, the energy spacing is on the order of $\hbar \omega/k_B T \sim 10^{-5}$. As such, we can approximate the sum in Eq. (\ref{eq:PartPlus}) as an integral,
\begin{equation}
	Z^+ \approx \frac{1}{2 \lambda} \int_{0}^{\infty} \mathrm{d}x \exp{\left(- \beta m c^2 \left[ \sqrt{x + 1} - 1\right]\right)},
\end{equation}
where $x \equiv 2 \lambda (n+1)$. Making the further substitution $u \equiv \sqrt{x + 1}$ this integral simplifies to, 
\begin{equation}
	Z^+ \approx \frac{1}{\lambda} \ex{\beta m c^2} \int_{1}^{\infty} \mathrm{d}u \, u \ex{- \beta m c^2 u},
\end{equation} 
which evaluates to,
\begin{equation}
	Z^+ \approx \frac{1}{\lambda} \left[\frac{1}{\left(\beta m c^2\right)^2} + \frac{1}{\beta m c^2}\right].
\end{equation} 
We can repeat an identical process to determine $Z^-$. The full partition function is then,
\begin{equation}
	\label{eq:DiracPartFcn}
	Z \approx \frac{1}{\hbar^2 \omega^2} \left(\frac{1}{\beta^2 m c^2} + \frac{1}{\beta}\right)^2.
\end{equation}

\begin{figure*}
	\centering
	\subfigure[]{
		\includegraphics[width=.46\textwidth]{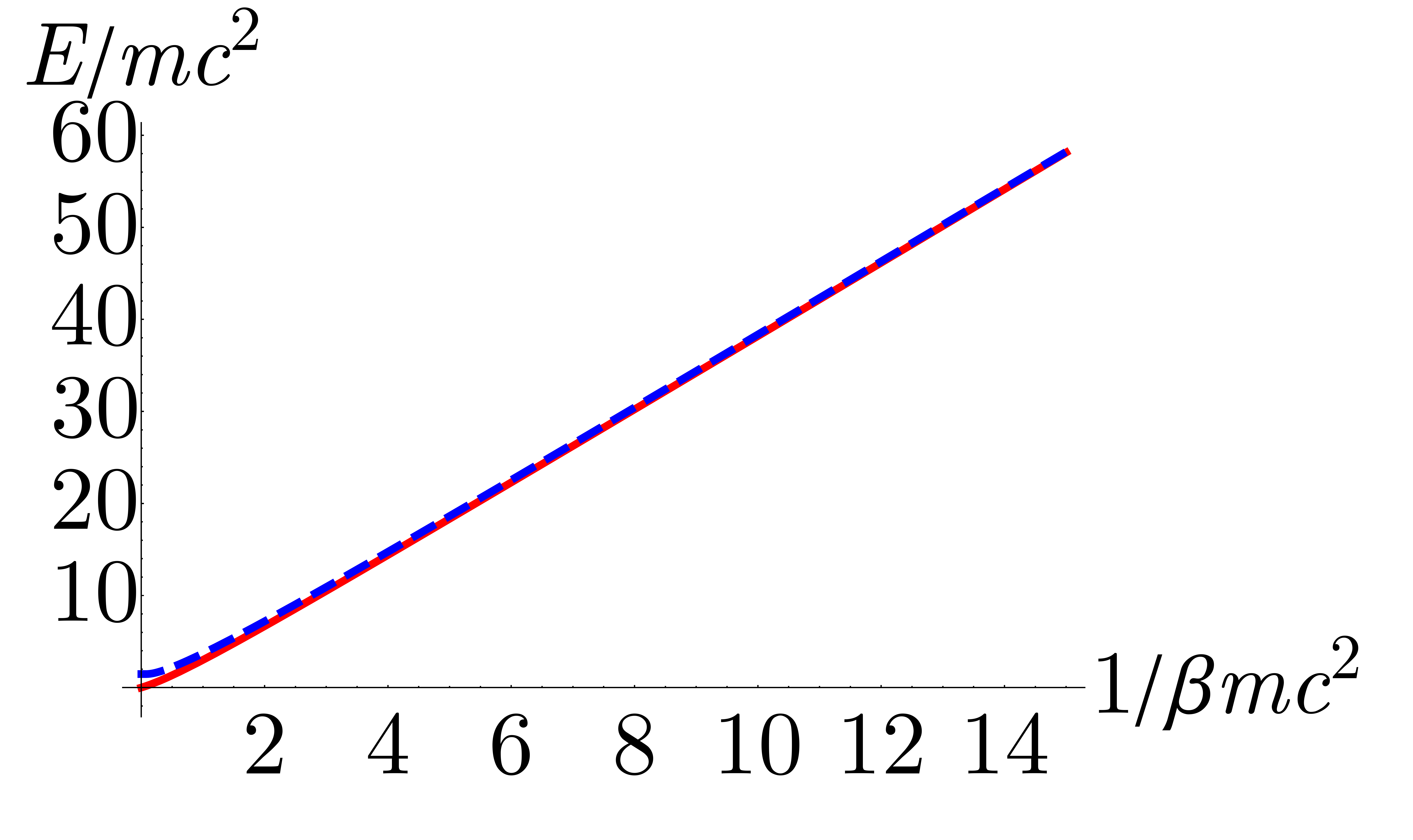}
	}
	\subfigure[]{
		\includegraphics[width=.46\textwidth]{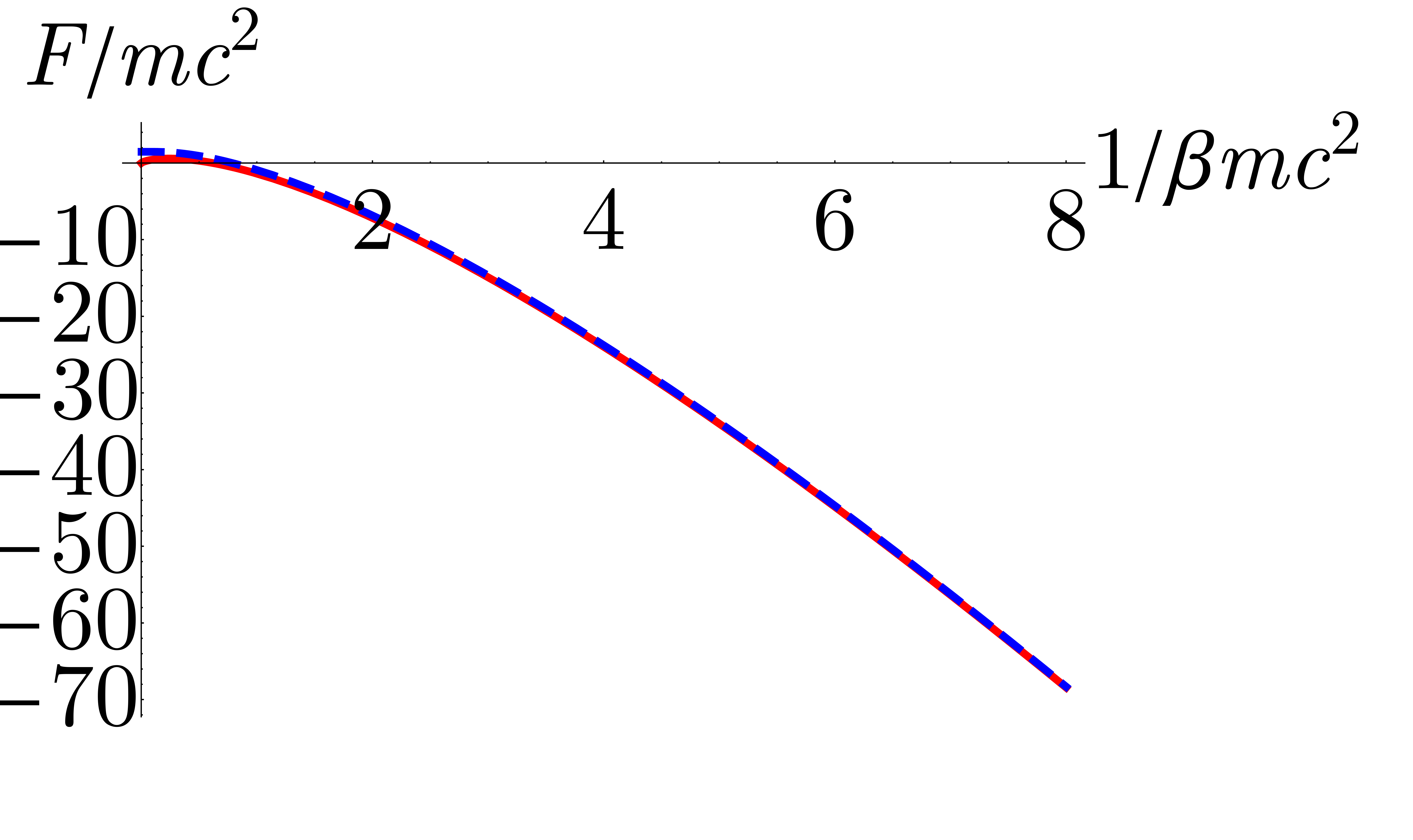}
	}
	\subfigure[]{
		\includegraphics[width=.46\textwidth]{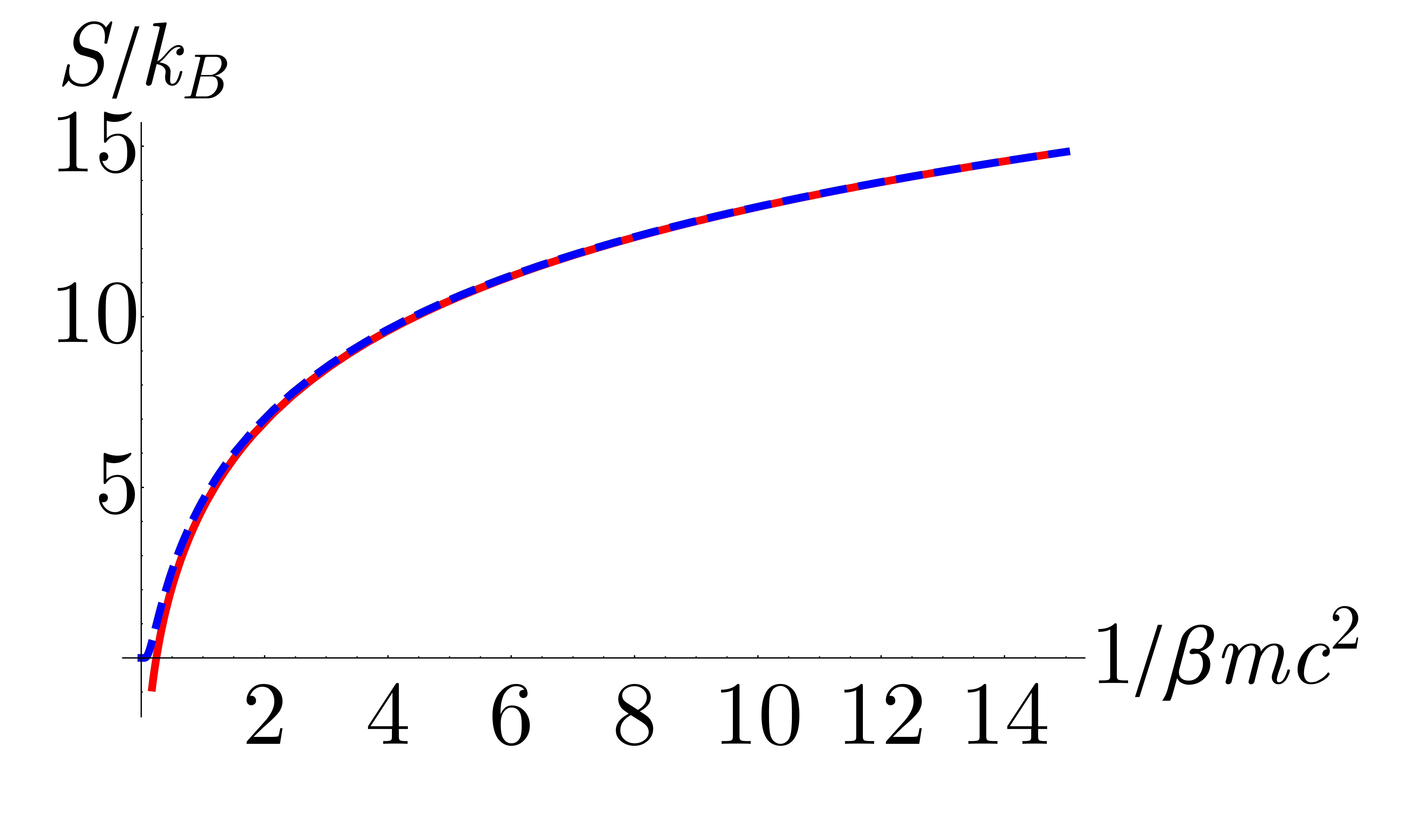}
	}
	\subfigure[]{
		\includegraphics[width=.46\textwidth]{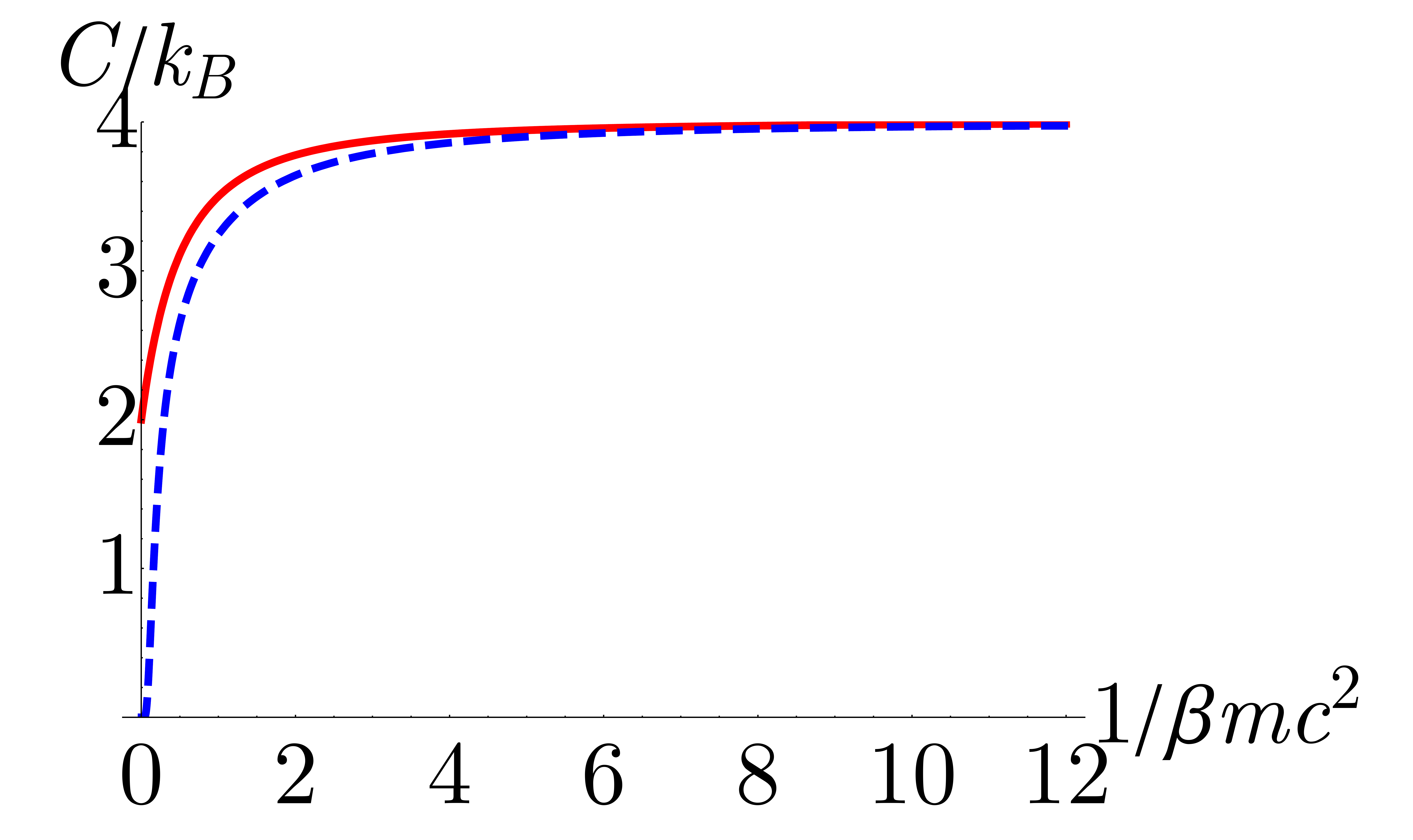}
	}
	\caption{\label{fig:ThermoQuant}Equilibrium (a) internal energy, (b) free energy, (c) entropy, and (d) heat capacity for the one-dimensional Dirac oscillator using numerical methods (blue, dashed) and the continuum approximation (red, solid). Parameters are set such that $\lambda = 1$.}
\end{figure*}

Combining Eqs. (\ref{eq:DiracPartFcn}) and (\ref{eq:thermofunc}) we find,
\begin{equation}
	\label{eq:IntThermoQuant}
	\begin{split}
		& \qquad \quad \,\,\, E = 2 \frac{2 + \beta m c^2}{\beta \left(1/\beta + \beta m c^2\right)}, \qquad \qquad \qquad \qquad \,
		F = -\frac{2}{\beta} \ln\left(\frac{1 + \beta m c^2}{\hbar \omega \beta^2 m c^2}\right), \\
		& S = 2 k_B + \frac{2 k_B}{1 + \beta m c^2} + 2 k_B \ln\left(\frac{1 + \beta m c^2}{\hbar \omega \beta^2 m c^2}\right), \qquad
		C = 2 k_B \frac{2 + \beta m c^2\left(4 + \beta m c^2\right)}{\left(1 + \beta m c^2\right)^2}.
	\end{split}
\end{equation}       
We note that the high temperature behavior of the thermodynamic quantities given in Eq. (\ref{eq:IntThermoQuant}) matches that found in Ref. \cite{Pacheco2003}. To determine the range of validity of the continuum approximation, in Fig. \ref{fig:ThermoQuant} we compare the analytical expressions found in Eq. (\ref{eq:IntThermoQuant}) to those found by calculating the partition function sum numerically, truncating after the first $15,000$ terms. We see that the continuum approximation matches with the numerical results very well outside of the low-temperature limit $\beta \rightarrow \infty$. We also note that in the continuum approximation the internal energy and heat capacity are independent of the frequency, indicating that system is behaving as an ideal gas \cite{Callen}. Similar results were found in Ref. \cite{Munoz2012} for the thermodynamics of a particle in a box obeying Dirac dynamics.      

\section{Endoreversible Relativistic Quantum Otto Engine}
\label{sec:4}

Just as for classical heat engines, the maximum efficiency of a quantum heat engine is given by the Carnot efficiency \cite{Deffner2019book}. However, achieving this efficiency requires infinitely long, quasistatic strokes, leading to vanishing power output. First introduced by Curzon and Ahlborn, the \textit{efficiency at maximum power} (EMP) provides a more practically useful metric of heat engine performance \cite{Curzon1975}. The EMP is found by first maximizing the power with respect to one of the system parameters, and then determining the efficiency at that power output. Using the framework of \textit{endoreversible thermodynamics} \cite{Curzon1975, Rubin1979, Hoffmann1997}, in which the working medium is assumed to be in a state of local equilibrium at all times, but with dynamics that occur quickly enough such that full equilibrium with the thermal reservoirs is not achieved, Curzon and Ahlborn found the EMP of a Carnot cycle to be \cite{Curzon1975},
\begin{equation}
	\label{eq:CAeff}
	\eta_{\mathrm{CA}} = 1 - \sqrt{\frac{T_c}{T_h}},
\end{equation}  
where $T_c$ ($T_h$) is the cold (hot) reservoir temperature. 

Numerous works have shown that the performance of a quantum Otto engine, including the EMP, depends on the nature of the working medium \cite{Uzdin2014, Pena2014, Zhang2014, Zheng2015, Jaramillo2016, Pena2017, Huang2017, Li2018, Kloc2019, YungerHalpern2019, Pena2019, Myers2020, Watanabe2020, Pena2020, Smith2020, Myers2021, Myers2021Sym}. In Ref. \cite{Deffner2018} an endoreversible Otto engine with a working medium of a single particle in a harmonic potential obeying Schr\"{o}dinger dynamics was found to have an EMP that exceeds the Curzon-Ahlborn efficiency (\ref{eq:CAeff}). To determine whether relativistic effects will further enhance or hinder the performance of such an engine we consider an endoreversible Otto engine with a working medium of a single particle in a one-dimensional relativistic oscillator potential. For ease of comparison to the non-relativistic case, we closely follow the analysis laid out in Ref. \cite{Deffner2018}.  

\begin{figure}
	\centering
	\includegraphics[width=.7\textwidth]{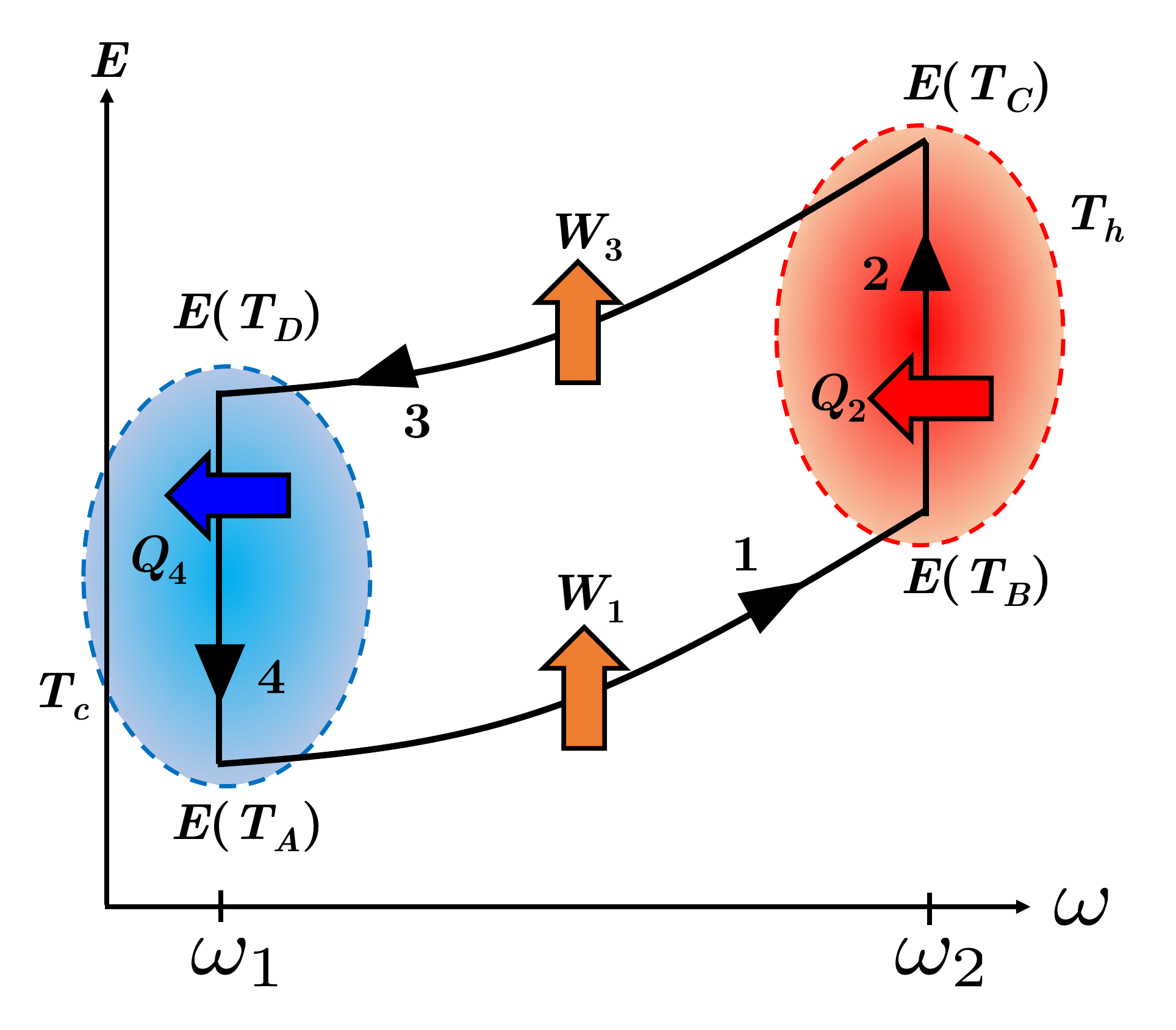}
	\caption{\label{fig:cycle} Energy frequency diagram of the endoreversible quantum Otto cycle.}
\end{figure}

The quantum Otto cycle, depicted in Fig. \ref{fig:cycle}, consists of four strokes:  \\

(1) \textit{Isentropic compression} \\

During this stroke the frequency of the relativistic oscillator is increased while the working medium remains isolated from the environment, ensuring that the entropy of the system remains constant. We can thus identify the change in internal energy during this stroke as work,
\begin{equation}
	\label{eq:Wcomp}
	W_{\mathrm{comp}} = E(T_B, \omega_2) - E(T_A, \omega_1).
\end{equation} 

(2) \textit{Isochoric heating} \\

During this stroke the frequency of the relativistic oscillator is held constant while the working medium exchanges heat with the hot reservoir. In accordance with the core assumption of endoreversible thermodynamics we assume that during this stroke the working medium does not fully equilibrate with the environment. We can identify the change in internal energy during this stroke as heat,
\begin{equation}
	\label{eq:Qh}
	Q_{h} = E(T_C, \omega_2) - E(T_B, \omega_2).
\end{equation}
We can determine the change in temperature of the working medium during its finite-time contact with the reservoir using Fourier' law \cite{Callen},
\begin{equation}
	\label{eq:Fouriers}
	\frac{dT}{dt} = -\alpha_h (T(t)-T_h),
\end{equation}
where $\alpha_h$ is a constant determined by the heat capacity and thermal conductivity of the working medium. Note that here we apply the linear Newtonian expression for heat conduction. We note that, in general, this linear approximation for diffusion is incompatible with special relativity, as it assumes an infinite speed for heat propagation \cite{Chester1963, Eckert1971, Ali2005}. For a relativistic open quantum system interacting with its environment the causality of diffusion is maintained if the condition $D \ll \hbar/(4m)$ is satisfied, where $D$ is the diffusion coefficient \cite{Cabrera2016}. For the following analysis we assume that the effective mean free path for the single particle working medium interacting with the thermal baths is very small, such that this condition is maintained. Solving Eq. (\ref{eq:Fouriers}) yields,
\begin{equation}
	\label{eq:fh}
	T_C - T_h  = (T_B - T_h) \ex{- \alpha_h \tau_h},
\end{equation} 
where $\tau_h$ is the duration of the heating stroke. 

(3) \textit{Isentropic expansion} \\

In complete analogy to the compression stroke, during the expansion stroke the frequency of the relativistic oscillator is decreased back to its original value while the working medium remains isolated from the environment. The work done during this stroke is then, 
\begin{equation}
	\label{eq:Wexp}
	W_{\mathrm{exp}} = E(T_D, \omega_1) - E(T_C, \omega_2).
\end{equation} 

(4) \textit{Isochoric cooling} \\

In the last stroke the working medium is brought into contact with the cold reservoir while the frequency is held constant. The heat exchanged with the cold reservoir is given by,
\begin{equation}
	Q_{c} = E(T_A, \omega_1) - E(T_D, \omega_1).
\end{equation}
As before, the temperature change can again be determined by solving Fourier's law,
\begin{equation}
	\label{eq:fc}
	T_A - T_c  = (T_D - T_c) \ex{- \alpha_c \tau_c},
\end{equation}
where $\tau_c$ is the duration of the cooling stroke.

The efficiency of the engine is given by the ratio of the total work and the heat exchanged with the hot reservoir,
\begin{equation}
	\label{eq:eff}
	\eta = -\frac{W_{\mathrm{comp}}+W_{\mathrm{exp}}}{Q_h},
\end{equation}
and the power output by the ratio of the total work to the cycle duration,
\begin{equation}
	\label{eq:P}
	P = -\frac{W_{\mathrm{comp}}+W_{\mathrm{exp}}}{\gamma (\tau_h + \tau_c)},
\end{equation}
with $\gamma$ serving as a multiplicative factor that implicitly incorporates the duration of the isentropic strokes \cite{Deffner2018}. 

Combining Eqs. (\ref{eq:IntThermoQuant}), (\ref{eq:Wcomp}), (\ref{eq:Wexp}), and (\ref{eq:Qh}) with (\ref{eq:eff}) and (\ref{eq:P}) yields complicated expressions for the efficiency and power in terms of the temperatures at each corner of the cycle, $T_A, T_B, T_C,$ and $T_D$. Ultimately, we want to instead express the efficiency and power in terms of the experimentally controllable parameters of the system, namely the initial and final oscillator frequencies and hot and cold reservoir temperatures. To do so, we can use the fact that the compression stroke satisfies the isentropic condition $dS = 0$. Noting that $S \equiv S(T, \omega)$, we can expand $dS$ as,
\begin{equation}
	dS = \left(\frac{\pd S}{\pd T} \right)_{\omega} dT + \left(\frac{\pd S}{\pd \omega} \right)_{T} d\omega = 0.
\end{equation}
Taking the appropriate partial derivatives of $S$ using Eq. (\ref{eq:IntThermoQuant}) and rearranging we arrive at,
\begin{equation}
	\frac{d \omega}{d\mathcal{T}} = \frac{\omega}{\mathcal{T}} \left[ \frac{1 + 2\mathcal{T} (2+\mathcal{T})}{(1 + \mathcal{T})^2} \right],
\end{equation}          
where $\mathcal{T} = k_B T/m c^2$ is the rescaled temperature. Separating variable and integrating we arrive at relationship between the initial and final frequencies and temperatures of the compression stroke,
\begin{equation}
	\label{eq:ABIsen}
	\frac{\omega_1}{\omega_2} = \frac{\mathcal{T}_A (1+ \mathcal{T}_A)}{\mathcal{T}_B (1+ \mathcal{T}_B)} \left\{ \frac{\mathrm{exp} \left(1/[1+\mathcal{T}_B]\right)}{\mathrm{exp} \left(1/[1+\mathcal{T}_A]\right)} \right\}.
\end{equation} 

At this point, let us take a moment to consider the three energy scales in the current analysis, namely $\hbar \omega$, $k_B T$, and $m c^2$. In Section \ref{sec:2} we noted that the Dirac oscillator spectrum reduces to that of the non-relativistic quantum harmonic oscillator in the limit $\hbar \omega \ll mc^2$. Conversely, the relativistic behavior is most apparent in the limit $\hbar \omega \gg mc^2$. With the introduction of thermal environments, we want to ensure the relativistic nature of the system does not change significantly during its interaction with the reservoirs. Thus in the relativistic case we assume that $k_B T \gg mc^2$ and in the non-relativistic case that $k_B T \ll mc^2$. In both the relativistic and non-relativistic limits, we assume that the condition of small energy level spacing, $\hbar \omega \ll k_B T$, still holds, such that the continuum approximation used to determine the partition function remains valid. Summarizing these relationships, we have in the relativistic limit that $k_B T \gg \hbar \omega \gg mc^2$ and in the non-relativistic limit that $m c^2 \gg k_B T \gg \hbar \omega$.

Assuming now the high-temperature, relativistic limit where $k_B T \gg mc^2$ we have $\mathcal{T}_A \gg 1$ and $\mathcal{T}_B \gg 1$. Consequently, the exponential factors in Eq. (\ref{eq:ABIsen}) reduce to one and the equation simplifies to,
\begin{equation}
	\frac{\omega_1}{\omega_2} = \left(\frac{\mathcal{T}_A}{\mathcal{T}_B}\right)^2.
\end{equation}
Defining the compression ratio, $\kappa = \omega_1/\omega_2$, we can rewrite this expression as,
\begin{equation}
	\label{eq:RelRelAB}
	\frac{\mathcal{T}_A}{\mathcal{T}_B} = \sqrt{\kappa}.
\end{equation}

\begin{figure}
	\centering
	\includegraphics[width=.72\textwidth]{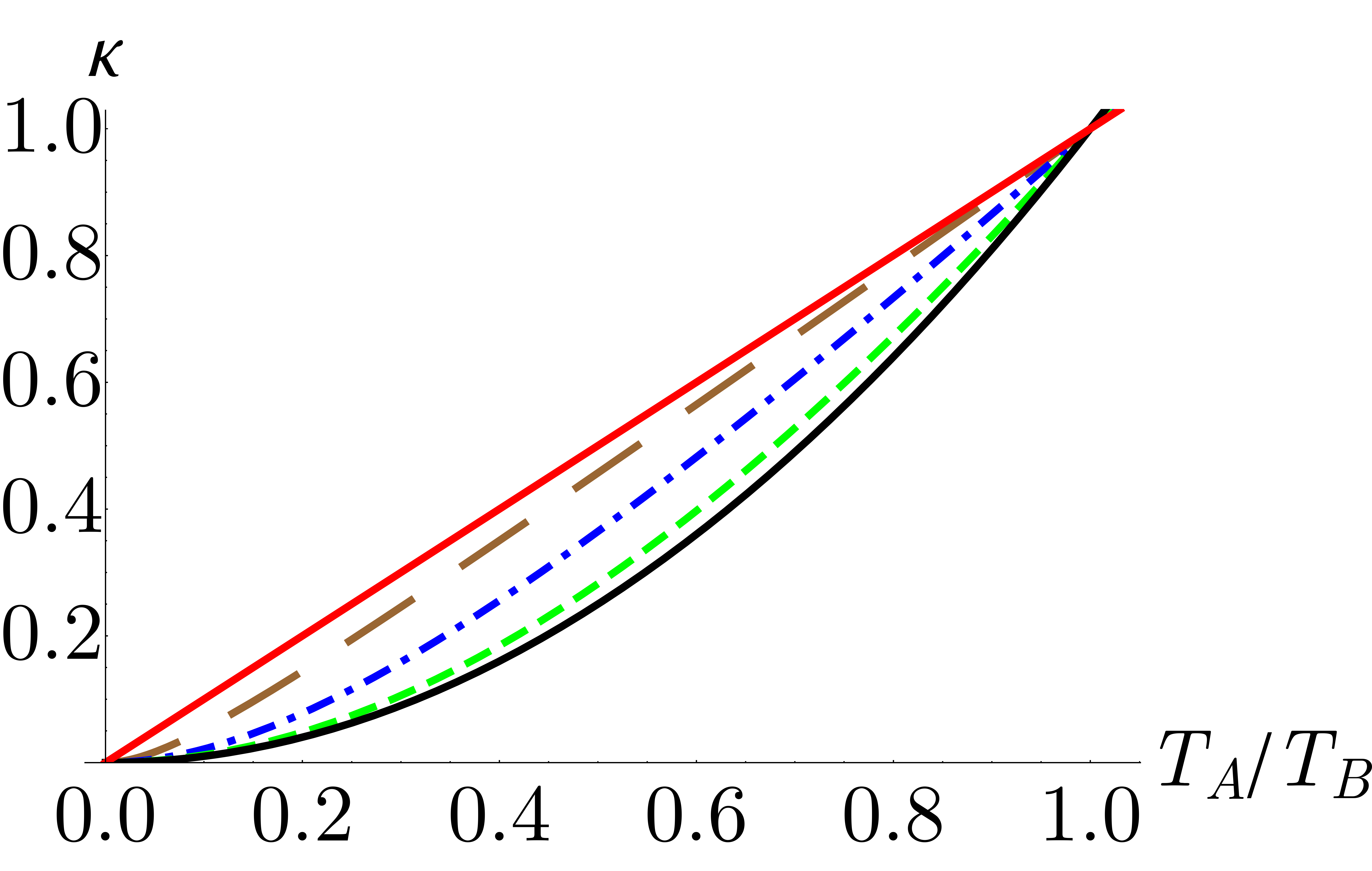}
	\caption{\label{fig:limitplot} Compression ratio as a function of the bath temperature ratio for $T_A = 0.05$ (brown, long dashed), $T_A = 0.25$ (blue, dot-dashed), and $T_A = 1$ (green, short dashed). The high temperature limit of $T_A/T_B = \sqrt{\kappa}$ (black, lower solid) and low temperature limit of $T_A/T_B = \kappa$ (red, upper solid) are given for comparison.} 
\end{figure}

In the low-temperature, non-relativistic limit where $k_B T \ll mc^2$ we will instead have $\mathcal{T}_A \ll 1$ and $\mathcal{T}_B \ll 1$. In this limit the exponential term in Eq. (\ref{eq:ABIsen}) will again reduce to one (as now $1 + \mathcal{T}_A \approx 1 + \mathcal{T}_B \approx 1$) and the full expression simplifies to,
\begin{equation}
	\label{eq:lowlimit}
	\frac{\omega_1}{\omega_2} = \frac{\mathcal{T}_A}{\mathcal{T}_B}.
\end{equation}
Rewriting Eq. (\ref{eq:lowlimit}) in terms of the compression ratio we have,
\begin{equation}
	\label{eq:NonRelRelAB}
	\frac{\mathcal{T}_A}{\mathcal{T}_B} = \kappa,
\end{equation}
which we recognize as the temperature-frequency relation that satisfies the isentropic condition for a harmonic Otto cycle obeying Schr\"{o}dinger dynamics \cite{Deffner2018, Myers2021}. The behavior of Eq. (\ref{eq:ABIsen}) is illustrated graphically in Fig. \ref{fig:limitplot}. We see that as the temperature of the baths increase, the behavior of the isentropic relation varies smoothly from the low-temperature behavior of Eq. (\ref{eq:NonRelRelAB}) to the high-temperature behavior of Eq. (\ref{eq:RelRelAB}).   

We can repeat this process for the expansion stroke, for which we find the relation,
\begin{equation}
	\label{eq:RelRelCD}
	\frac{\mathcal{T}_D}{\mathcal{T}_C} = \sqrt{\kappa},
\end{equation}
in the high-temperature, relativistic limit and,
\begin{equation}
	\label{eq:NonRelRelCD}
	\frac{\mathcal{T}_D}{\mathcal{T}_C} = \kappa,
\end{equation}
in the low temperature, non-relativistic limit. 

\subsection{Efficiency}

\begin{figure}
	\centering
	\includegraphics[width=.7\textwidth]{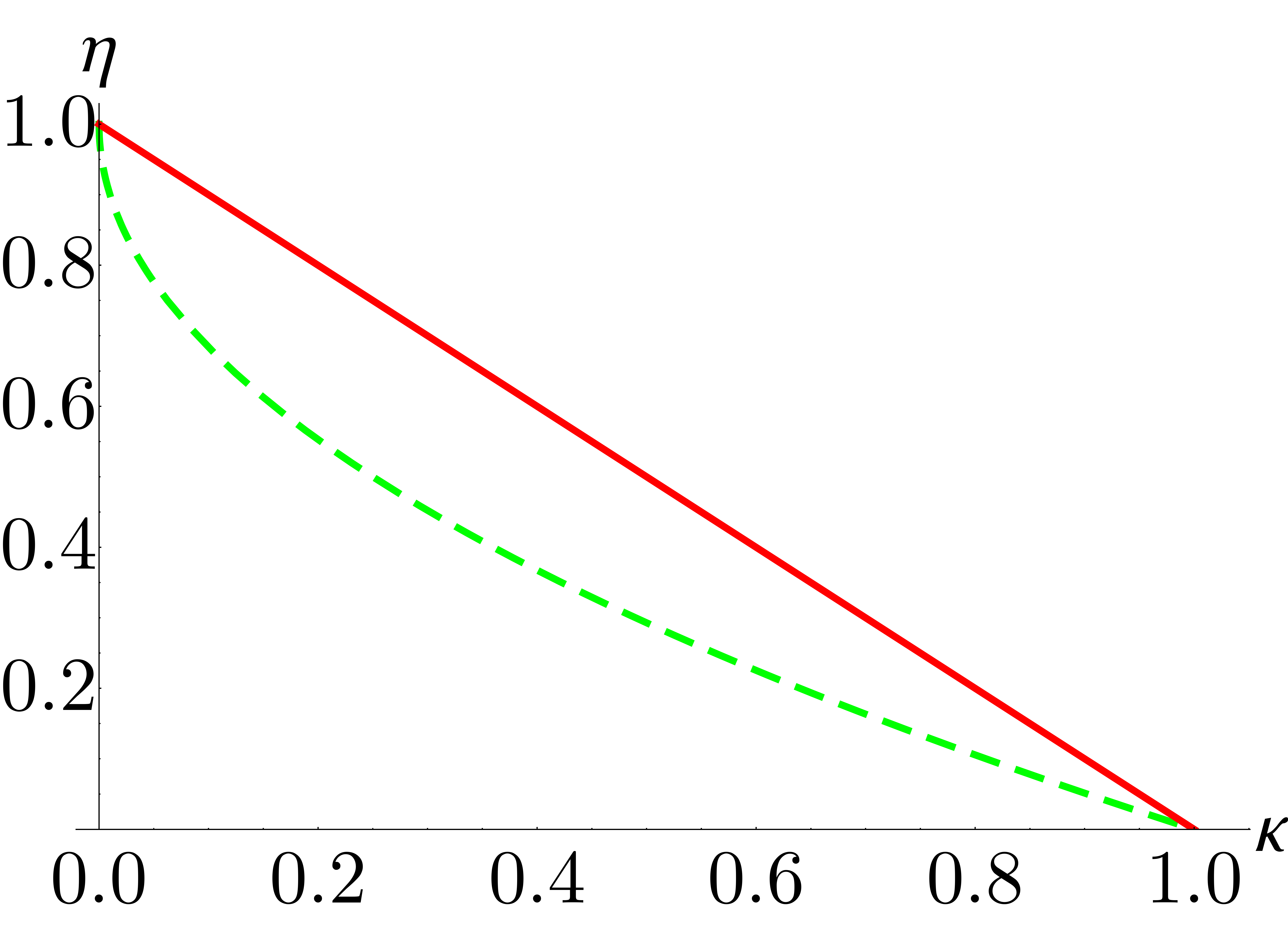}
	\caption{\label{fig:effplot} Efficiency as a function of the compression ratio for an endoreversible quantum engine with a relativistic oscillator working medium in the relativistic limit (green, dashed) and non-relativistic limit (red, solid).}
\end{figure} 

\begin{figure*}
	\centering
	\subfigure[]{
		\includegraphics[width=.65\textwidth]{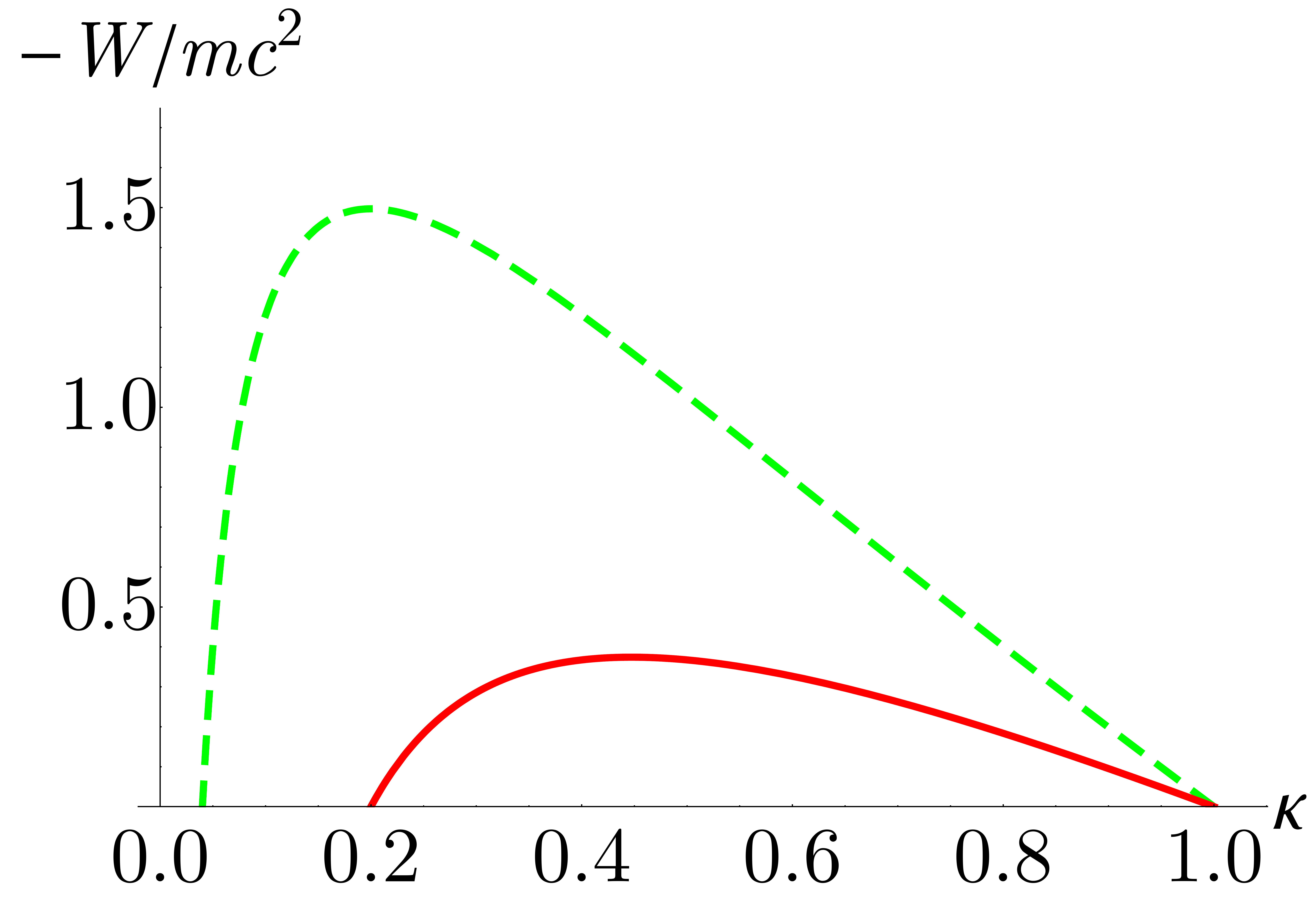}
	}
	\subfigure[]{
		\includegraphics[width=.65\textwidth]{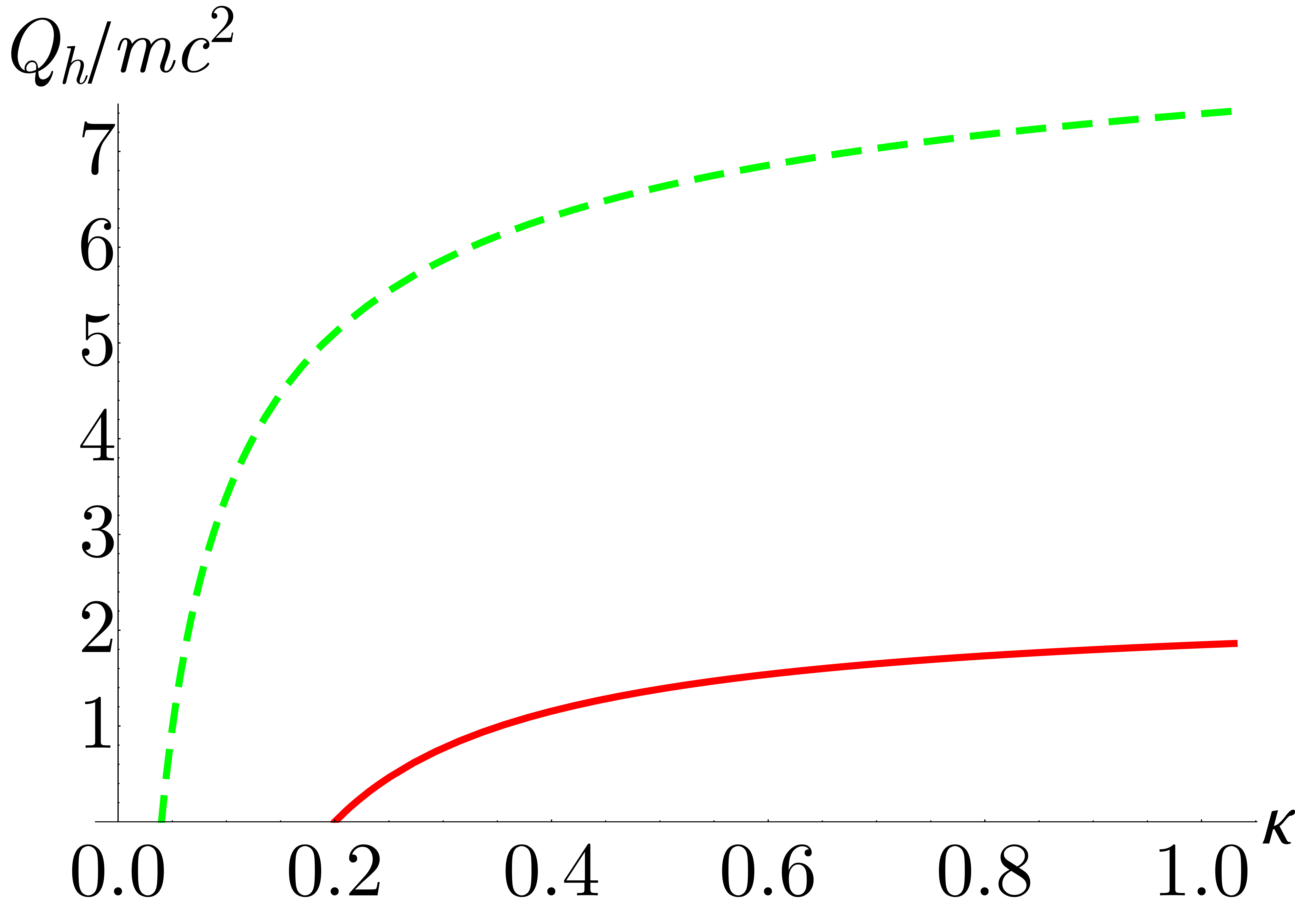}
	}
	\caption{\label{fig:WorkHeatplot}Total work (a) and heat exchanged with the hot bath (b) as a function of $\kappa$ in the relativistic (green, dashed) and non-relativistic (red, solid) limits. Parameters are $\alpha_c = \alpha_h = 1$, $\tau_c = \tau_h = 0.5$ and $T_c/T_h = 1/5$.}
\end{figure*} 

Let us first consider the relativistic limit. Combining Eqs. (\ref{eq:eff}) with (\ref{eq:fh}), (\ref{eq:fc}), (\ref{eq:RelRelAB}), and (\ref{eq:RelRelCD}) and taking the high-temperature limit we arrive at a much simplified expression for the efficiency,
\begin{equation}
	\label{eq:effHigh}
	\eta_{\mathrm{rel}} = 1 - \sqrt{\kappa}.
\end{equation}
Repeating this process for the non-relativistic, low-temperature limit (now using Eqs. (\ref{eq:NonRelRelAB}) and (\ref{eq:NonRelRelCD}) rather than Eqs. (\ref{eq:RelRelAB}) and (\ref{eq:RelRelCD})) we find,
\begin{equation}
	\label{eq:effLow}
	\eta_{\mathrm{nonrel}} = 1 - \kappa.
\end{equation}
We note that this efficiency is identical to the efficiency found for an endoreversible harmonic quantum Otto engine obeying Schr\"{o}dinger dynamics \cite{Deffner2018}, as we would expect in the non-relativistic limit. The only difference between the relativistic and non-relativistic efficiencies is the presence of the square root of the compression ratio. This follows from the fact that the Dirac oscillator potential is linear in the frequency, while the harmonic oscillator potential is quadratic in the frequency. 

In Fig. \ref{fig:effplot} we plot the relativistic and non-relativistic limits of the efficiency as a function of the compression ratio. We see that the relativistic working medium displays significantly reduced efficiency when compared to the non-relativistic working medium. We can understand this intuitively by noting the frequency plays the role of the inverse volume. The linear frequency scaling of the Dirac oscillator, which leads to the square root of the compression ratio, means that the relativistic system is working with an effectively reduced volume. This additional restriction arises as the allowed state transitions, and accompanying changes in momentum, for the relativistic system are constrained by the light cone \cite{Deffner2015}. 

\begin{figure}
	\centering
	\includegraphics[width=.7\textwidth]{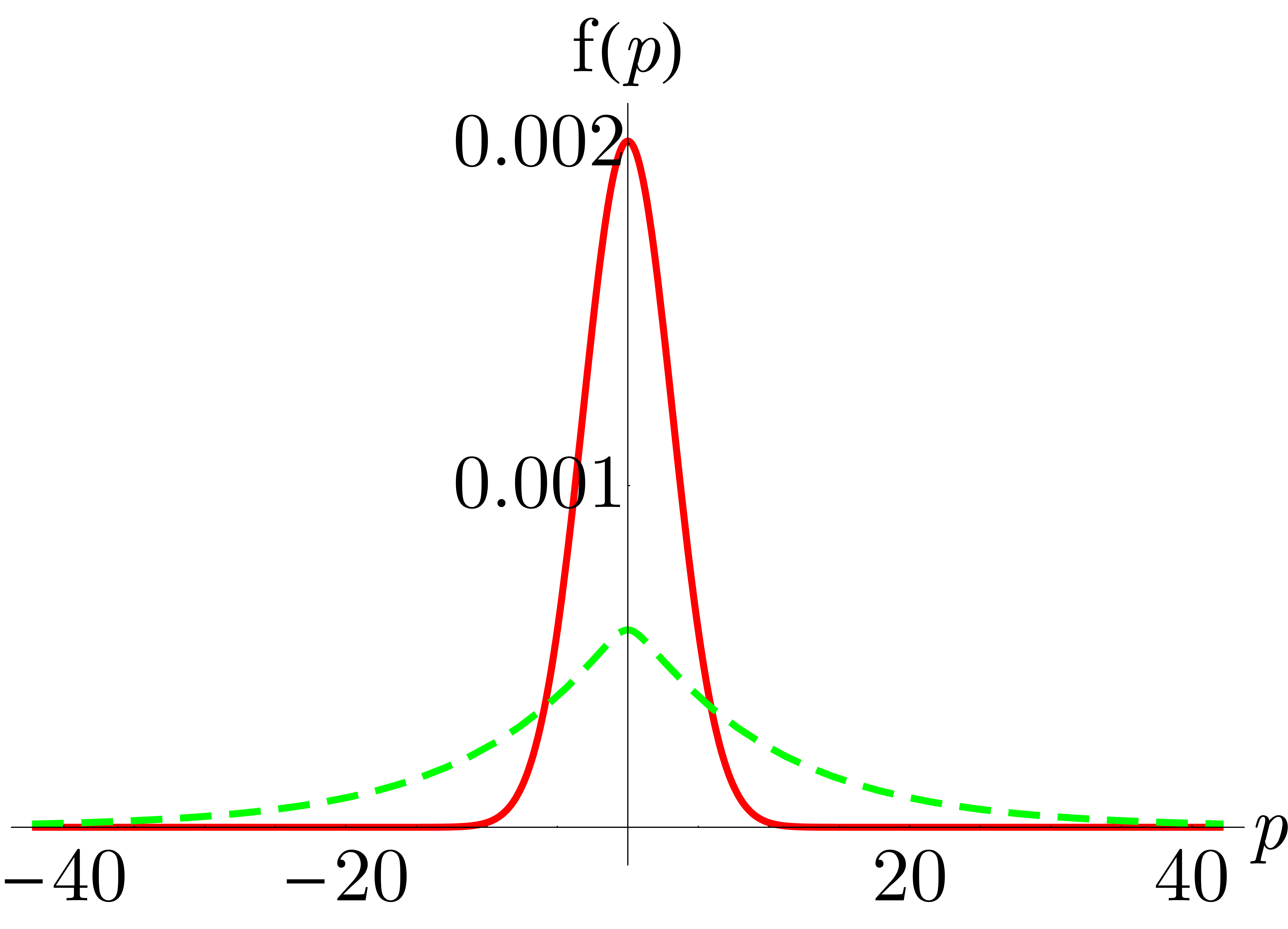}
	\caption{\label{fig:MJdist} Maxwell-Boltzmann thermal momentum distribution for particles obeying Schr{\"o}dinger dynamics (red, solid) in comparison to the Maxwell-J{\"u}ttner thermal momentum distribution for particles obeying Dirac dynamics (green, dashed). Parameters are $m = c = k_B = 1$.}
\end{figure}                

For further insight, we plot the total work and the heat exchanged with the hot reservoir as a function of the compression ratio in Fig. \ref{fig:WorkHeatplot} (where we use the convention that work extracted from the system is negative). We see that for all $\kappa$ the relativistic medium extracts more work. Additionally, the compression ratio at which the maximum work is achieved is smaller than for the non-relativistic medium. The enhanced work output has two main contributions. The first is the additional factor of two in the relativistic internal energy that accounts for the existence of the negative energy solutions. The second contribution arises from the fact that, due to the linear rather than quadratic dependence on momentum, each degree of freedom for an ultra-relativistic gas contributes twice the amount to the internal energy as in the classical, non-relativistic limit \cite{Pathria2011}. This ``relativistic equipartition" is derived from the fact that the thermal momentum distribution of particles obeying Dirac dynamics follows the Maxwell-J{\"u}ttner distribution rather than the typical Maxwell-Boltzmann distribution found for particles obeying Schr{\"o}dinger dynamics \cite{Juttner1911, Deffner2015}. The Maxwell-J{\"u}ttner distribution, illustrated in Fig. \ref{fig:MJdist}, is broader than the Gaussian Maxwell-Boltzmann distribution, leading to a higher probability of large momentum values. Intuitively, this leads to increased work extraction, as large values of work are associated with large changes in momentum. This behavior can be seen in the work distribution for a Dirac particle which is strongly peaked around large values of work \cite{Deffner2015}. The effectively reduced volume of the relativistic oscillator also shifts the maximum work value to a smaller compression ratio, as larger variations in $\omega$ are required to produce the same change in effective volume.

From Fig. \ref{fig:WorkHeatplot}a we also see that the positive work condition is violated at the point that $\kappa$ is equal to the square of the bath temperature ratio for the relativistic medium and when $\kappa$ is equal to the bare bath temperature ratio for the relativistic medium. Comparing this to Eqs. \eqref{eq:effHigh} and \eqref{eq:effLow} we see that this corresponds to the point at which the efficiency reaches the Carnot efficiency. Thus for both mediums the Carnot efficiency is only achievable in the limit of vanishing work output.      

Examining Fig. \ref{fig:WorkHeatplot}b we see that, due to the same contributions to the internal energy, the relativistic medium also exchanges more heat with the hot reservoir. As such, the relativistic contributions to the heat and work cancel out when determining the efficiency. This leads to the overall lower efficiency for the relativistic medium, due to the effectively reduced compression ratio. On the other hand, we note that the reduced compression ratio of the relativistic medium means that a wider range of frequencies meet the ``positive work condition" ($W < 0$ and $Q_h > 0$) under which the cycle operates as an engine.  

\subsection{Efficiency at Maximum Power}
	
To determine the power output of the relativistic engine we repeat the steps used to find the total work that we applied for the efficiency, using the isentropic conditions and solving Fourier's law to express the temperature at each corner of the cycle in terms of the compression ratio and bath temperatures. The algebra can be considerably simplified by first taking the corresponding high or low temperature limit of the internal energy before applying Eq. (\ref{eq:P}). For the low-temperature, non-relativistic medium, the power output is given by, 
\begin{equation}
	\label{eq:PLow}
	P_{\mathrm{nonrel}} = \frac{2 \left(\kappa-1\right) k_B \left(T_c - \kappa T_h \right)}{\gamma \kappa (\tau_c + \tau_h)} \frac{\sinh \left(\alpha_c \tau_c/2\right) \sinh \left(\alpha_h \tau_h/2\right)}{\sinh \left([\alpha_c \tau_c + \alpha_h \tau_h]/2\right)}.
\end{equation}
We note that this is identical to the power output of an endoreversible harmonic Otto engine operating in the ``classical regime" where $\hbar \omega_2/k_B T_c \ll 1$ as found in Ref. \cite{Deffner2018}. This is expected, as the classical regime condition is equivalent to the assumption of very small energy spacing that we made to convert the partition function sum to an integral. Similarly, for the high-temperature, relativistic medium the power is given by,
\begin{equation}
	\label{eq:PHigh}
	P_{\mathrm{rel}} = \frac{8 \left(\sqrt{\kappa }-1\right) k_B \left(T_c - \sqrt{\kappa}T_h \right)}{\gamma \sqrt{\kappa} (\tau_c + \tau_h)} \frac{\sinh \left(\alpha_c \tau_c/2\right) \sinh \left(\alpha_h \tau_h/2\right)}{\sinh \left([\alpha_c \tau_c + \alpha_h \tau_h]/2\right)}.
\end{equation}
As in the low-temperature limit, the relativistic power is similar in form to that of a classical harmonic Otto engine, but with the compression ratio replaced by the reduced ratio, $\sqrt{\kappa}$, along with an additional overall factor of four arising from the negative energy solutions and the linear momentum dependence.          

\begin{figure}
	\centering
	\includegraphics[width=.7\textwidth]{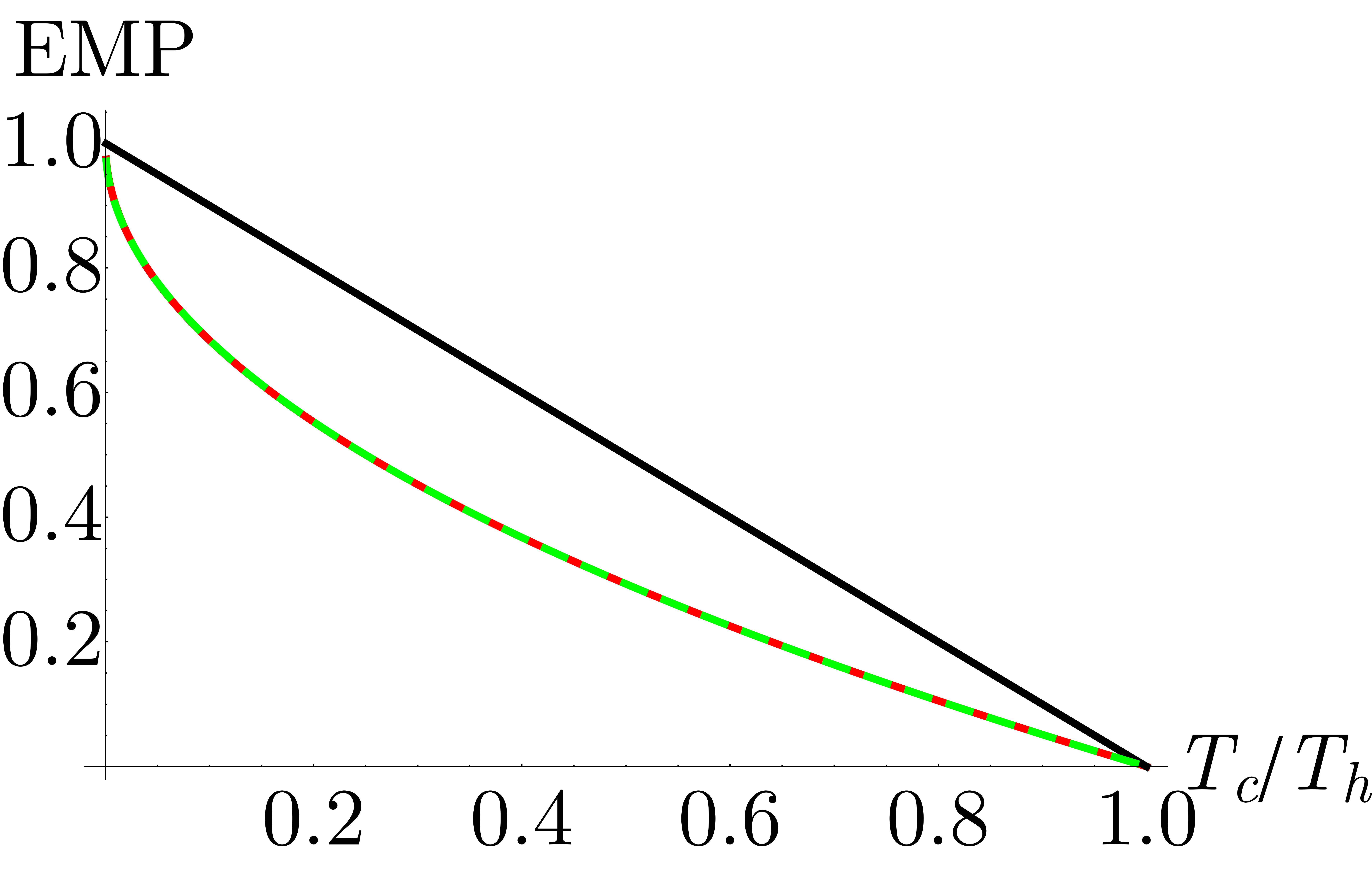}
	\caption{\label{fig:EMPplot} EMP as a function of the bath temperature ratios for a relativistic (green, dashed), and a non-relativistic (red, bottom solid) working medium. The Carnot efficiency (black, top solid) is given for comparison. Note that the relativistic EMP and non-relativistic EMP are identical to the Curzon-Ahlborn efficiency. Parameters are $\gamma = \alpha_c = \alpha_h = 1$ and $\tau_c = \tau_h = 0.5$.}
\end{figure}

It is straightforward to show that Eqs. (\ref{eq:PLow}) and (\ref{eq:PHigh}) are maximized when $\kappa = \sqrt{T_c/T_h}$ and $\kappa = T_c/T_h$, respectively. We can now determine the EMP by plugging these values of $\kappa$ into Eq. (\ref{eq:effHigh}) (for the relativistic medium) and Eq. (\ref{eq:effLow}) (for the non-relativistic medium). This yields,
\begin{equation}
	\eta_{\mathrm{rel}} = 1 - \sqrt{\kappa}  = 1 - \sqrt{T_c/T_h} \quad \text{and} \quad \eta_{\mathrm{nonrel}} = 1 - \kappa = 1 - \sqrt{T_c/T_h},
\end{equation} 
for the relativistic and non-relativistic working mediums, respectively. We see directly that, while they display different efficiencies and work extraction, the EMP of both working mediums are identical to the Curzon-Ahlborn efficiency, Eq. (\ref{eq:CAeff}). The EMP as a function of the bath temperature ratios is plotted in Fig. \ref{fig:EMPplot}, along with the Carnot efficiency for comparison.     

\section{Discussion}       
\label{sec:5}

\subsection{Potential Experimental Implementations}

While a widely-studied theoretical system, it is only recently that Dirac oscillator dynamics have been demonstrated experimentally using an array of microwave resonators \cite{Franco2013}. Experimental realizations of Dirac dynamics have also been implemented using trapped ions \cite{Gerritsma2010} and in Dirac materials such as graphene \cite{Novoselov2005}. Such systems present the opportunity to study  the microscopic dynamics of naturally-occurring relativistic systems, such as black holes and cosmic jets \cite{Meyer2018}, at laboratory-accessible energies. Furthermore, Dirac materials have direct practical applications, such as the implementation of field effect transistors in graphene using relativistic charge carriers \cite{Novoselov2004}. From both a practical and scientific standpoint it is important that we understand the thermodynamic behavior of such systems. Heat engines provide a well established framework for doing so. With this in mind, we propose how three experimental systems might be generalized to construct a relativistic quantum Otto engine.

\subsubsection{Trapped ions:}

Trapped ions have played a prominent role in quantum thermodynamics as a paradigmatic system for implementing nanoscale harmonic heat engines \cite{Abah2012, Rossnagel2016, Dawkins2018}. As such, trapped ions are a logical first place to look when envisioning an experimental implementation of a relativistic oscillator heat engine. In Ref. \cite{Lamata2007} an experimental simulation of free particle Dirac dynamics was proposed (and later implemented in Ref. \cite{Gerritsma2010}) by mapping the internal levels of the atom to the components of the Dirac bispinor. These internal levels are then coupled to the motional degrees of freedom of the trapped atom through a Jaynes-Cummings interaction implemented with a laser field. In Ref. \cite{Bermudez2007} this work was extended to further show that the dynamics of a Dirac oscillator could be directly mapped onto the Jaynes-Cummings model. Using this framework, the re-scaled oscillator frequency, $\hbar \omega/m c^2$, is controlled through the excitation coupling strengths and trap frequency \cite{Bermudez2007}. The isentropic strokes of the cycle could be implemented through modification of these parameters. The isochoric strokes could be implemented in the typical manner for single ion engines, using de-tuned laser beams to excite or de-excite the ionic vibrational modes \cite{Abah2012}. 

\subsubsection{Dirac materials:} 

The relativistic dispersion relations observed for charge carriers in graphene has led to numerous proposals centered around using the 2D material as a test bed for observing relativistic phenomena. A recent experimental proposal outlined how a graphene chip could be used to observe relativistic Brownian motion \cite{Pototsky2012}. In this framework, the graphene sheet is placed on a series of electrodes with alternating constant potentials. By tuning the distance between electrodes the total potential experienced by the charge carriers can be modulated at will \cite{Pototsky2012}. In this manner a relativistic oscillator potential could be constructed and varied in strength in order to implement the isentropic strokes of the cycle. The isochoric strokes could then be implemented by coupling the graphene sheet alternately to a high and low temperature thermal environment, or through the application of a global noisy elctrostatic force \cite{Martinez2016}.

\subsubsection{Microwave resonators:}

Proposed as an alternative for simulating the relativistic dynamics observed in Dirac materials \cite{Sadurn2010}, microwave resonator arrays provided the first experimental demonstration of the Dirac oscillator \cite{Franco2013}. Based on the correspondence between the Dirac oscillator Hamiltonian and a distorted tight-binding model \cite{Sadurn2010}, the experimental system is constructed from a chain of resonator disks in dimeric configuration placed between two metallic plates \cite{Franco2013}. By controlling the coupling between sites through the interdisk distance the resonances of the microwave system can be tuned to yield the Dirac oscillator spectrum. It is worth noting that this method is only capable of producing a finite portion of the oscillator spectrum, as each pair of disks corresponds to a single energy state \cite{Franco2013}. A gap in the spectrum that can be interpreted as an effective particle ``mass" can be introduced by ensuring that the distance between each disk in a dimer pair is less than the maximal distance between dimers \cite{Franco2013}. While the isentropic strokes of an Otto cycle can be straightforwardly executed in this framework by varying the coupling between sites to change the oscillator frequency, the isochoric strokes are more difficult to carry out as it is not immediately clear how to introduce a method of thermalization. However, it may be possible to implement a thermal state by using a noisy field to excite the resonance modes of the disks.                 

\subsection{Concluding Remarks}

In this work we have examined an endoreversible quantum Otto engine with a working medium of a single particle in a relativistic oscillator potential. We have shown that, while the relativistic working medium extracts more work than a non-relativistic medium, it also draws more heat from the hot reservoir and displays universally reduced efficiency in comparison to the non-relativistic medium. We determined analytical expressions for the power in both the relativistic and non-relativistic limits, and found that both relativistic and non-relativistic working mediums display efficiency at maximum power equivalent to the Curzon-Ahlborn efficiency, within the parameter regime under which the continuum approximation for the partition function sum is valid. 

While our analysis has focused on the Otto cycle, consisting of isentropic and isochoric strokes, the fact that both heat and work are impacted by relativistic features indicates that other stroke types during which both quantities are exchanged, such as isothermal and isobaric strokes \cite{Quan2009}, may also display relativistic effects. This suggests that the performance of cycles that incorporate these strokes, such as Carnot, Stirling, Brayton, or Diesel, will also be altered by relativistic behavior. Future work could also extend the current analysis beyond the endoreversible regime, to fully non-equilibrium, time-dependent dynamics. This opens up the intriguing possibility of optimizing the engine performance by applying relativistic shortcuts to adiabaticity. Enhancing quantum engine performance through shortcuts to adiabaticity has been studied extensively for the case of Schr{\"o}dinger dynamics \cite{Abah2017, Abah2018, Abah2019, Beau2016, Campo2014, Funo2019, Bonanca2019, Baris2019, Dann2020, Li2018}, but to our knowledge this idea remains totally unexplored for the case of Dirac or Klein-Gordon dynamics. However, it has been demonstrated that fast forward \cite{Deffner2015NJP}, inverse engineering \cite{Song2017}, and time-rescaling \cite{Roychowdhury2021} shortcut methods can be generalized to relativistic quantum systems.

While there remains much to do in order to develop an engine that could power the starship \textit{Enterprise}, we have demonstrated that, in principle, the extraction of work from relativistic effects is possible.     

\invisiblesection{Acknowledgments}
\section*{Acknowledgments}
It is a pleasure to thank Francisco J. Pe{\~n}a for enlightening conversations on the topic of relativistic quantum engines. N. M. gratefully acknowledges support from Harry Shaw of NASA Goddard Space Flight Center and Kenneth Cohen of Peraton. This material is based upon work supported by the U.S. Department of Energy, Office of Science, Office of Workforce Development for Teachers and Scientists, Office of Science Graduate Student Research (SCGSR) program. The SCGSR program is administered by the Oak Ridge Institute for Science and Education for the DOE under contract number DE‐SC0014664. O.A. acknowledges support from the UK EPSRC EP/S02994X/1. S.D. acknowledges support from the U.S. National Science Foundation under Grant No. DMR-2010127.  

\appendix

\invisiblesection{Appendix}
\section*{Appendix: Relativistic Partition Function Derivation}
\label{Appendix}
\renewcommand{\theequation}{A.\arabic{equation}}
\setcounter{equation}{0}  

In this appendix we provide the full details of the derivation of the canonical equilibrium state for a relativistic oscillator by maximizing the state multiplicity using the method of Lagrange multipliers. 

Consider a system of $k$ Dirac oscillators with $N$ energy quanta that can be distributed among them. For a fixed total energy the number of ways to distribute the $N$ total energy quanta is,
\begin{equation}
	\Omega = \frac{N!}{n_1!n_2!...n_k!},
\end{equation}         
where $n_i$ is the occupation of the $i$th state, either positive or negative. Assuming that $\Omega$ is large, we can apply Stirling's approximation,
\begin{equation}
	\label{eq:logMult}
	\ln{\Omega} \approx N \ln{N} - \sum_{i = 1}^{k} n_i \ln{n_i}.
\end{equation}
To determine the equilibrium occupation, we want to maximize $\Omega$ with respect to $n_i$. Due to the stability of the Dirac sea, we note that that the contributions to the multiplicity from the positive and negative energy states must be independently maximized. We can do this using the method of Lagrange multipliers,
\begin{equation}
	\label{eq:LagrangeMult}
	\delta \left[\ln{\Omega} + \alpha^+ \sum_{i^+ = 1}^{k^+} n_i + \alpha^- \sum_{i^- = 1}^{k^-} n_i + \beta^+\sum_{i+}^{k^+}n_i \epsilon_i - \beta^-\sum_{i-}^{k^-}n_i \epsilon_i \right] = 0,
\end{equation} 
where $\alpha^+$, $\alpha^-$, $\beta^+$, and $\beta^-$ are Lagrange multipliers. The constraints,
\begin{equation}
	N^+ = \sum_{i^+}^{k^+}n_i \quad \text{and} \quad N^- = \sum_{i^-}^{k^-}n_i, 
\end{equation}
arise from normalization of the positive and negative energy states, respectively, and,
\begin{equation}
	E^+ = \sum_{i+}^{k^+}n_i \epsilon_i \quad \text{and} \quad E^- = -\sum_{i-}^{k^-}n_i \epsilon_i,
\end{equation}
from energy conservation. Plugging in Eq. (\ref{eq:logMult}) and separating the first sum into sums over the positive and negative occupancies Eq. (\ref{eq:LagrangeMult}) becomes,
\begin{equation}
	\begin{split}
		\delta \Bigg[ & \sum_{i+}^{k^+} \left( -n_i \ln{n_i} + \alpha^+ n_i + \beta^+ n_i \epsilon_i \right) \\
		&\qquad +\sum_{i-}^{k^-} \left( -n_i \ln{n_i} + \alpha^- n_i - \beta^- n_i \epsilon_i \right) \Bigg] = 0.
	\end{split}
\end{equation}    
Since positive and negative energy states corresponding to the same energy level can not be simultaneously occupied, we require that each term in both sums must be separately zero,
\begin{equation}
	\label{eq:LagrangePos}
	\left(- \ln{n_i} + \alpha^+ + \beta^+ \epsilon_i \right) \delta n_i = 0,
\end{equation}
and,
\begin{equation}
	\label{eq:LagrangeNeg}
	\left(- \ln{n_i} + \alpha^- - \beta^- \epsilon_i \right) \delta n_i = 0.
\end{equation}
Solving Eq. (\ref{eq:LagrangePos}) and applying the normalization constraint we have,
\begin{equation}
	n_i^+ = \frac{1}{Z^+} \ex{- \beta^+ \epsilon_i},
\end{equation}
where,
\begin{equation}
	Z^+ = \sum_{i+} \ex{- \beta^+ \epsilon_i}.
\end{equation}
Similarly, Eq. (\ref{eq:LagrangeNeg}) yields,
\begin{equation}
	n_i^- = \frac{1}{Z^-} \ex{\beta^- \epsilon_i},
\end{equation}
where,
\begin{equation}
	Z^- = \sum_{i-} \ex{\beta^- \epsilon_i}.
\end{equation}

\section*{References}

\bibliographystyle{iopart-num}
\bibliography{diracRefs} 

\end{document}